\documentclass[onecolumn,showpacs,floatfix,superscriptaddress,longbibliography]{revtex4-1}
\usepackage{mathrsfs}
\usepackage{amssymb, amsbsy, amsmath, latexsym, dsfont, array, layout,mathrsfs,color,ulem,bm,upgreek}
\usepackage[colorlinks,linkcolor=blue,anchorcolor=red,citecolor=blue]{hyperref}
\usepackage{multirow}
\usepackage{float}
\usepackage{threeparttable}
\usepackage{geometry}
\usepackage{natbib}

\usepackage{graphicx}
\usepackage{xspace, stmaryrd, ulem}
\definecolor{delete}{rgb}{1.0, 0.0, 0.0}
\definecolor{edit}{rgb}{0.0, 0.0, 0.9}
\definecolor{comment}{rgb}{0.9, 0.0, 0.0}

\begin{document}
\title{An experimental proposal certification for any three-qubit generalized Greenberger-Horne-Zeilinger states based on the fine-grained steering inequality}

\author{Zhi-Hao Bian}\email{zhihaobian@jiangnan.edu.cn}

\author{Jia-Qi Sun}

\author{Yi Shen}
\affiliation{School of Science, Jiangnan University, Wuxi 214122, China}

\begin{abstract}
Multi-party quantum steering is an important concept in quantum information theory and quantum mechanics, typically related to quantum entanglement and quantum nonlocality. It enables precise manipulation of large quantum systems, which is essential for large-scale quantum computing, simulations, and quantum communication. Recently, a quantum steering certification for any three-qubit generalized Greenberger-Horne-Zeilinger (GGHZ) states based on the fine-grained steering inequality was proved [Quantum Studies: Mathematics and Foundations, 2022, 9(2): 175-198]. Here we provide an experimental proposal to prepare the GGHZ states in photon system. The measurement observalbes in each party can be realized by different polarization optical elements. By choosing the angles of the waveplates, our experiment proposal can observe the maximum quantum violation for any three-qubit GGHZ states. Our proposal can be easily extended to high-dimensional qubits and multi-photon GHZ states, which provides a method to study the complex multi-party quantum protocols.
\end{abstract}

\maketitle

\section{Introduction}
Quantum entanglement between spatially separated parties is one of the most fascinating phenomena in quantum physics. It plays a crucial role as a fundamental resource in quantum information processing. As an essential element of quantum nonlocality~\cite{LCB+14,P12,hirsch2013genuine}, quantum entanglement is widely used in applications such as quantum dense coding~\cite{horodecki2009quantum,mattle1996dense,chen2021orbital}, quantum teleportation~\cite{bouwmeester1997experimental,lipka2021catalytic}, and quantum distillation~\cite{regula2021fundamental,fang2020no,kalb2017entanglement}. Verifying and characterizing specific types of quantum entanglement is vital for advancing quantum information technologies. A bipartite quantum state is considered entangled if and only if it is not separable. Similarly, a pure multipartite quantum state is termed genuinely multipartite entangled if it cannot be separated in any bipartition~\cite{G09}. Due to its intricate nature, detecting genuine multipartite entanglement experimentally is highly challenging. In an ideal scenario, assuming full trust in the measurement devices, one could directly access the quantum state and perform quantum state tomography to reconstruct the density matrix, which contains all the state’s information. However, as the system's dimension increases, the number of required projective measurements grows exponentially, making state tomography impractical, especially for high-dimensional systems and limited state copies. Moreover, in real-world applications, the devices are often untrusted, further complicating the verification process. In such situations, quantum state tomography may not be the optimal approach for verifying entanglement. The question naturally arises as to whether it is possible to certify the presence of genuine multipartite entanglement in a scenario where some of the observers' devices are not fully characterized~\cite{bowles2018device,bancal2014device,zwerger2019device}.

EPR steering, first introduced by Schrödinger~\cite{schrodinger1935discussion}, serves as a quintessential example of quantum nonlocality, enabling the detection of bipartite and multipartite states in partially device-independent systems. Later, Wiseman et al. subsequently formalized this phenomenon in the context of an operational task~\cite{wiseman2007steering}. The concept of quantum steering can be briefly introduced as follows: Consider two parties, Alice and Bob, who share a bipartite quantum state (e.g., an entangled state) and each have access to one of the subsystems. Let’s assume Alice performs a measurement on her subsystem, which may affect the state of Bob’s subsystem. The goal of quantum steering is to understand how Alice’s measurement influences Bob’s system. The state is classified as "steerable" if the corresponding post-measurement state for Bob’s subsystem cannot be accounted for by any local hidden state (LHS) model. Typically, if Alice aims to steer Bob, she can send him the measurement outputs, while Bob, in turn, performs a quantum measurement on the collapsed reduced density matrix. This approach allows one party to conduct classical measurements, thereby enabling the possibility that one of the measurement devices may be untrusted.

Due to the above characteristics, quantum steering is typically associated with the secret key rate in one-sided device-independent quantum key distribution~\cite{branciard2012one,gheorghiu2017rigidity}. In addition, quantum steering offers advantages in self-testing pure entangled states~\cite{goswami2018one,vsupic2016self} and distinguishing quantum states~\cite{bian2020experimental}. Currently, quantum steering inequalities have been proposed and experimentally demonstrated in various systems~\cite{cavalcanti2015analog,orieux2018experimental,wollmann2016observation}. However, the theoretical and experimental studies on quantum steering are mostly focused on bipartite systems, with relatively few investigations in multipartite systems. To tackle this challenge, device-independent (DI) witnesses for genuine multipartite entanglement can be utilized~\cite{nagata2002configuration,bancal2014device,augusiak2019bell,maity2020detection}. These DI witnesses depend solely on measurement statistics, eliminating the need for detailed characterization of the experimental devices. This makes them a robust tool for entanglement verification in practical scenarios.

In Ref.~\cite{gupta2022all}, the authors extend the Greenberger-Horne-Zeilinger (GHZ) theorem to any three-qubit generalized Greenberger-Horne-Zeilinger (GGHZ) state and present an "all-versus-nothing" proof of genuine tripartite EPR steering for all three-qubit pure GGHZ states in both one-sided and two-sided device-independent scenarios. They also propose a two-sided device-independent (2SDI) tripartite steering inequality, which generalizes the fine-grained steering inequality (FGSI) from the bipartite scenario. Here, we present an experimental proposal for entanglement certification of any three-qubit generalized GGHZ states based on a linear optical system, demonstrating that for any GGHZ state, the FGSI can achieve the maximal violation. Our article is organized as follows. In Section~\ref{sec:2}, we give a brief introduction of the tripartite steering inequality based on fine-grained uncertainty relation. Section~\ref{sec:3} is the experimental proposal about preparation of GGHZ states to certificate the FGSI. In Section~\ref{sec:4}, we discuss and summarize our results.

\section{Tripartite steering inequality based on fine-grained uncertainty relation}\label{sec:2}

Multipartite quantum steering is a generalization of quantum steering to systems involving more than two parties~\cite{cavalcanti2011unified,cavalcanti2016quantum}. It refers to the ability of one party to steer or influence the state of a distant, entangled quantum system shared between several parties by performing local measurements on their subsystem. In the case of multipartite quantum steering, the parties involved can steer the states of others in a way that reflects quantum nonlocality, and it is a feature that arises due to the entanglement between all parties in the system. For simplicity, we give a detailed breakdown of the mathematical structure and the quantification of tripartite quantum steering.

Suppose Alice, Bob, and Charlie share a tripartite an unknown quantum system. The state of this system is described by a density matrix $\rho_{ABC}$ , where $A$ refers to Alice's subsystem, $B$ refers to Bob's subsystem, and $C$ refers to Charlie's subsystem respectively. Now first consider Alice will perform local measurements $A_{i}$ on her subsystem based on a dichotomic order $i\in \{0, 1\} $, and her measurement outcome will influence the conditional states of Bob and Charlie. The measurement is described by a set of positive operator-valued measurements (POVM) $\{M_{a|A_i}\}$ with $M_{a|A_i}\geq 0$ and $\sum_a M_{a|A_i}=\mathbb{I}$, where $a_i$ represents the measurement outcomes. The quantum state of Bob and Charlie's subsystems is then updated conditionally on Alice's outcomes, which are known as conditional states $\sigma_{a|A_i}^{BC}$. The unnormalized conditional state is given by tracing over Alice's subsystem, denoted as~\cite{cavalcanti2016quantum}
\begin{equation}
\sigma_{a|A_i}^{BC}=\text{Tr}_A[(M_{a|A_i}\otimes\mathbb{I}\otimes\mathbb{I})\rho_{ABC}].
\end{equation}
For quantum steering to occur, the conditional state must be non-separable, meaning it cannot be explained by any LHS model. The inseparability implies that Alice's measurement has steered Bob and Charlie's states in a manner that cannot be explained classically. The above scenario is called one-sided device-independent due to the POVM elements associated with Alice’s measurements are unknown.

In this article, we focus on the two-sided device-independent scenario. In this case, Bob performs local measurements $B_{j}$ on her subsystem based on a dichotomic order $j\in \{0, 1\} $. Similar to Alice, the measurement of Bob is described by a set of POVM $\{M_{b|B_j}\}$ with $M_{b|B_j}\geq 0$ and $\sum_b M_{b|B_j}=\mathbb{I}$, where $b_j$ represents the measurement outcomes and the POVM elements are unknown. When Alice obtains the measurement result $a$ and Bob obtains the measurement result $b$ simultaneously, the unnormalized conditional state of Charlie's subsystems is updated conditionally by~\cite{cavalcanti2011unified,cavalcanti2016quantum}
\begin{equation}
\sigma_{ab|A_iB_j}^{C}=\text{Tr}_{AB}[(M_{a|A_i}\otimes M_{b|B_j}\otimes\mathbb{I})\rho_{ABC}].
\end{equation}

If the conditional state $\sigma_{ab|A_iB_j}^{C}$ of Charlie is separable (i.e., it can be written as a product state), then the system can be explained by a LHS model, and no quantum steering is occurring. In the LHS model, each subsystem of the composite system performs measurements on their respective subsystems, and the outcomes are determined by a set of local hidden variables (LHVs) $\lambda$ that are present in the system but not directly observable. The corresponding hidden variable could represent, for example, a specific value that governs the outcome of measurements in that subsystem. The state of the overall system which is named the local hidden state, can be described by a probability distribution over the hidden variables, denoted by $\rho_\lambda$ with $\rho_\lambda\geq 0$ and $\text{Tr}\rho_\lambda=1$. The probability distribution $P(\lambda)$ is defined on the set of all possible hidden states, which represents the unknown or hidden information that determines the outcomes of measurements on the system and satisfies $P(\lambda)\geq 0$ and $\sum_\lambda P(\lambda)=1$. Thus, the conditional state $\sigma_{ab|A_iB_j}^{C}$ can be decomposed in the form~\cite{gupta2022all}
\begin{equation}
\sigma_{ab|A_iB_j}^{C}=\sum_\lambda P(\lambda)P(a|A_i,\lambda)P(b|B_j,\lambda)\rho^C_\lambda.
\end{equation}

Now consider Charlie performs a set of POVM $\{M_{c|C_k}\}$ with $M_{c|C_k}\geq 0$ and $\sum_c M_{c|C_k}=\mathbb{I}$ based on a dichotomic order $k\in \{0, 1\}$, where $c_k$ represents the measurement outcomes. Unlike the unknown measurements of Alice and Bob, Charlie performs quantum measurement on his subsystem. Then the measurement correlations of the tripartite system can be defined as a joint probability distribution $P(abc|A_iB_jC_k)=\text{Tr}[M_{c|C_k}\sigma_{ab|A_iB_j}^{C}]$. In the LHV-LHV-LHS model, this probability distribution can be written as
\begin{equation}
\label{CP}
P(abc|A_iB_jC_k)=\sum_\lambda P(\lambda)P(a|A_i,\lambda)P(b|B_j,\lambda)P(c|C_k,\rho^C_\lambda),
\end{equation}
where $P(a|A_i,\lambda)$ and $P(b|B_j,\lambda)$ represent the probability of obtaining the outcomes $a$ and $b$ when Alice and Bob apply $A_i$ and $B_j$ on each subsystem under the LHVs $\lambda$ respectively. In comparison, $P(c|C_k,\rho^C_\lambda)$ is the quantum probability of obtaining the outcome $c$ when Charlie performs the measurement on the LHS $\rho^C_\lambda$.

Essentially speaking, if the joint probability distribution can be decomposed in the form Eq.~(\ref{CP}), the shared state of system can be divided into three bi-separable terms since it contains no genuine entanglement~\cite{cavalcanti2015detection}. The first one implies that the assemblage updated conditionally on Alice's outcomes cannot be steered from Alice to Charlie, representing a lack of steerability in $A\rightarrow C$ direction. The second one denotes that the assemblage updated conditionally on Bob's outcomes cannot be steered from Bob to Charlie, representing a lack of steerability in $B\rightarrow C$ direction. The last one is the unsteerable assemblage from Alice and Bob to Charlie. Eq.~(\ref{CP}) can be viewed in terms of a hybrid LHS model, which highlights the non-steerability of the assemblage in different directions and provides a way to analyze its potential for demonstrating quantum correlations. If the measurement correlations can be expressed in the form of Eq.~(\ref{CP}), the system state does not demonstrate tripartite steering.

Next, we consider the conditional probability distribution in the hybrid LHS model, that is,
\begin{equation}
P(c|ab)=\frac{P(abc|A_iB_jC_k)}{P(ab|A_iB_j)}=\frac{\sum_\lambda P(\lambda)P(a|A_i,\lambda)P(b|B_j,\lambda)P(c|C_k,\rho^C_\lambda)}{P(ab|A_iB_j)}.
\end{equation}
By using the Cauchy-Schwarz Inequality $\sum_ix_iy_i\leq \text{max}\{x_i\}\sum_iy_i$, we can obtain
\begin{equation}
\begin{aligned}
P(c|ab)&\leq \text{max}\{P(c|C_k,\rho^C_\lambda)\}(\frac{\sum_\lambda P(\lambda)P(a|A_i,\lambda)P(b|B_j,\lambda)}{P(ab|A_iB_j)})  \\
&=\text{max}\{P(c|C_k,\rho^C_\lambda)\}=P(c|C_k,\rho^C_{\lambda_{\text{max}}}).
\end{aligned}
\end{equation}

Here we introduce the fine-grained uncertainty relation into the hybrid LHS model~\cite{goswami2018one,bian2020experimental,oppenheim2010uncertainty,bian2021experimental1}. The uncertainty relation is a fundamental concept in quantum mechanics that describes the inherent limits to how precisely certain pairs of physical properties, such as position and momentum, can be simultaneously measured or known~\cite{sanchez1998position}. Below is a brief overview of the fine-grained uncertainty relation: Assume Alice has a qubit, she receives a dichotomic order (denoted as $I$), which is either 0 or 1, with equal probability $P(I)=1/2$. If $I=0(1)$ Alice applies the $\sigma_x (\sigma_y)$ to the qubit, where $\sigma_i$ ($i=x,y,z$) is the Pauli matrix correspond to measurements of spin along the certain axis respectively. When $\sigma_x (\sigma_y)$ is applied, the measurement result $a$ is either 0 (spin-up) or 1 (spin-down), depending on the qubit's state. We define that Alice wins if she measures the spin-up outcome, then regardless of whether Alice receives $I=0$ or $I=1$, she aims to measure $a=0$. Therefore, the winning probability which depends on the outcome of the measurements satisfies
\begin{equation}
P=\sum_I p(I)p_{a=0}\leq P_{\text{max}},
\end{equation}
where $P_{\text{max}}$ is the maximum winning probability among all possible strategies which Alice can choose. It is straightforward to check that $P_{\text{max}}=\frac{1}{2}+\frac{1}{2\sqrt{2}}$ when the qubit state is the eigenstates of $\frac{\sigma_x+\sigma_y}{\sqrt{2}}$. Similarly, the maximum winning probability $P_{\text{max}}=\frac{1}{2}+\frac{1}{2\sqrt{2}}$ when Alice measures the spin-up outcome with the qubit state is the eigenstates of $\frac{\sigma_x-\sigma_y}{\sqrt{2}}$~\cite{oppenheim2010uncertainty}.

Under the fine-grained uncertainty relation, the sum of conditional probabilities satisfies~\cite{gupta2022all}
\begin{equation}
\label{FGI}
\begin{aligned}
S&=P(c_{C_0}|a_{A_0}b_{B_0})+P(c_{C_1}|a_{A_0}b_{B_1})+P(c_{C_0}|a_{A_1}b_{B_1})+P(c_{C_1}|a_{A_1}b_{B_0})\\
&\leq 2\ \text{max}\{P(c|\tilde{C_0},\rho^C_\lambda)+P(c|\tilde{C_1},\rho^C_\lambda)\},
\end{aligned}
\end{equation}
where $P(c|\tilde{C},\rho^C_\lambda)$ in the right-hand side represents the maximum possible winning probability for Charlie since the trusted party Charlie performs two arbitrary mutually unbiased qubit measurements. The measurements $\tilde{C_0}$ and $\tilde{C_1}$ are chosen to be maximally incompatible which range over all possible incompatible measurements. In quantum mechanics, the incompatible measurements cannot be simultaneously specified with exact certainty since they cannot be diagonalized simultaneously, which leads to the largest possible uncertainty in measurement outcomes, that is the reason that the right-hand side represents the maximum possible values that the conditional probabilities can achieve.

Eq.~(\ref{FGI}) is a fine-grained steering criterion satisfied by tripartite states. It shows that Alice and Bob can steer Bob if it violates for any combination of outcomes $\{a, b, c\}$. In this case, Alice and Bob try to maximize the right-hand side of Eq.~(\ref{FGI}) using the hybrid LHS model under the Alice-Bob’s strategy to cheat Charlie when Charlie’s particle is a qubit. Assume that Alice and Bob get the certain information of measurement before preparing the tripartite state, the maximum winning probability $P_{\text{max}}$ of Charlie is $\frac{1}{2}+\frac{1}{2\sqrt{2}}$ regardless of whether he measures the spin-up outcome or the spin-down outcome. Thus, the FGSI can be written as~\cite{gupta2022all}
\begin{equation}
S\leq 2[(\frac{1}{2}+\frac{1}{2\sqrt{2}})+(\frac{1}{2}+\frac{1}{2\sqrt{2}})]=2+\sqrt{2}.
\end{equation}

In another case, Alice and Bob obtain no prior knowledge of Charlie’s measurement at the beginning. To ensure to win the game, Charlie needs to scan all the possible measurement strategies and then maximize the winning probability in regard to all possible local hidden states. The average winning probability over all possible observables is proven to $3/4$, Therefore, the fine-grained steering inequality can be written as~\cite{gupta2022all}
\begin{equation}
S\leq 2(\frac{3}{4}+\frac{3}{4})=3.
\end{equation}

It is clearly that Eq.~(\ref{FGI}) has a set of different equivalent forms since $a, b, c \in \{0, 1\}$. As a demonstration, we can write
\begin{equation}
\label{FGS}
\begin{aligned}
S&=P(0_{C_0}|1_{A_0}1_{B_0})+P(0_{C_1}|0_{A_0}1_{B_1})+P(0_{C_0}|0_{A_1}1_{B_1})+P(0_{C_1}|0_{A_1}1_{B_0})\\
&\leq 2+\sqrt{2}  \     \   \   \   \   \   \ \ \ \    (C_k \   \text{is known})         \\
&\leq 3  \ \ \ \  \ \ \ \ \    \   \ \ \ \ \ \ \ \               (C_k \   \text{is unknown}).
\end{aligned}
\end{equation}
Quantum violation of the above inequality implies tripartite quantum steering in 2SDI scenario.

In the following we analyse the witness of the 2SDI FGSI for any pure state that belongs to the GGHZ class, the forms are given by
\begin{equation}
|\psi\rangle=\cos\theta|000\rangle+\sin\theta|111\rangle,   \  \ \ \ 0<\theta<\frac{\pi}{2}.
\end{equation}
It is easily to check that any pure GGHZ state $|\psi\rangle$ violates Eq.~(\ref{FGS}) to its algebraic maximum of 4, and the following choice of measurements are~\cite{gupta2022all}
\begin{equation}
\label{M}
\begin{aligned}
&A_0=\sigma_x;\ \ \ \ B_0=\sin 2\theta\sigma_x+\cos 2\theta\sigma_z;\ \ \ \ C_0=\sigma_x \\
&A_1=\sigma_y;\ \ \ \ B_1=\sin 2\theta\sigma_y+\cos 2\theta\sigma_z;\ \ \ \ C_1=\sigma_y.
\end{aligned}
\end{equation}

\section{Experimental proposal}\label{sec:3}
The entanglement certification for any three-qubit GGHZ states based on the FGSI contains the GGHZ states preparation and the measurement of three parties. Here we provide an experimental proposal by photons system. Photons, as carrier of qubits, possess unique properties that make them an excellent choice for quantum technologies~\cite{bian2015realization,bian2017experimental,xue2015experimental}. Photons can be measured with great accuracy in terms of polarization, position, momentum, etc~\cite{bian2020quantum,bian2020conserved}. Quantum measurement techniques such as polarization analysis and interference measurement are well-developed, enabling high-fidelity detection of entangled states~\cite{bian2021experimental2}. Compared to other systems (like superconducting qubits or ion traps), photons are easier to manipulate in experiments. Through standard optical components such as beam splitters, waveplates, fiber optics, and interferometers, one can precisely control the photon's polarization, phase, and other quantum properties. It is important that photons allow for the creation of multiple entangled photon pairs simultaneously. Using several photon sources and interferometers, it's possible to scale the system up to generate multi-photon GHZ states~\cite{hamel2014direct,bouwmeester1999observation,cao2022experimental,ren2023multipartite,guo2023experimental}. The scalability of optical components enables quantum computing and communication systems to support an increasing number of qubits. Photons can be easily extended to high-dimensional qubits, allowing for the entanglement of many photons. This is crucial for complex quantum algorithms and multi-party quantum protocols.

\begin{figure}[H]
\centering
\includegraphics[width=14 cm]{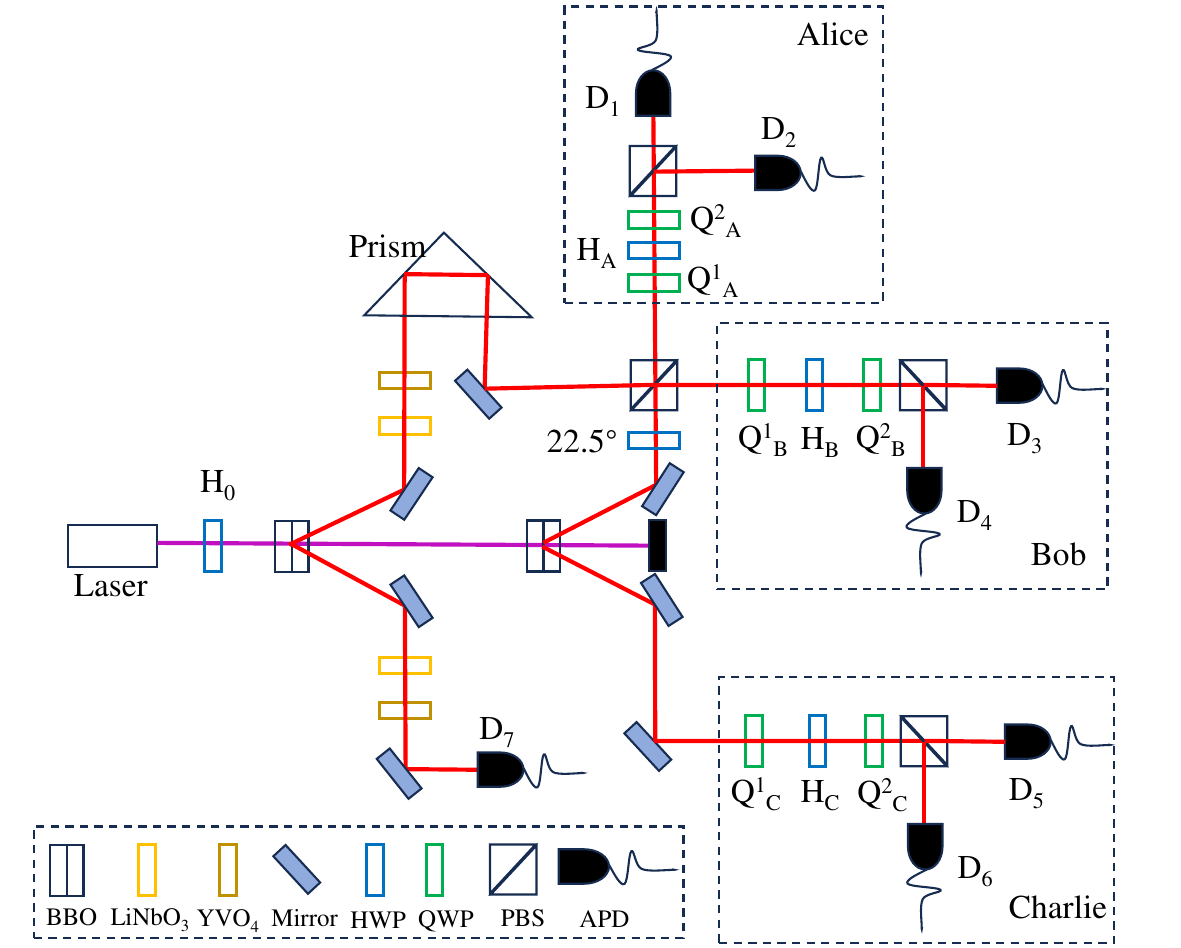}
\caption{Experimental proposal. The experimental setup can be divided into state preparation module and measurement module. Polarization encoded three-photon GGHZ states are produced by combining two pairs of
entangled photons generated from two $\beta$-BBO crystals pumped by a laser with the setting angle of the HWP $H_0=\theta/2$ applied to the pump light via type-I spontaneous parametric down-conversion process. LiNbO$_3$ and YVO$_4$ are used for spatial and temporal compensations between horizontal and vertical polarizations respectively. The observables of Alice, Bob and Charlie in the whole experiment are realized by of a combination of sandwich type QWP-HWP-QWP sequence and a following PBS. All the joint probabilities can be read out from the coincidence between certain APDs. Key to components: BBO, $\beta$-barium-borate crystal; HWP, half-wave plate; QWP, quarter-wave plate; PBS,polarizing beam splitter; APD, avalanche photodiodes.}
\label{fig:1}
\end{figure}

In the proposal, the computational basis states $|0\rangle$ and $|1\rangle$ are represented by the horizontal polarization $|H\rangle$ and vertical polarization $|V\rangle$ of a photon respectively, which is shown in Fig.~(\ref{fig:1}). Polarization encoded three-photon GHZ states are produced by combining two pairs of entangled photons generated from two back-to-back nonlinear $\beta$-barium-borate (BBO) crystals with two optical axes being vertical to each other pumped by a laser. After the process of spontaneous parametric down-conversion, two ultrabright EPR photon sources are used to generate the tripartite GHZ state~\cite{guo2023experimental}. Here an Hong-Ou-Mandel interferometer ensures photons from different EPR sources are indistinguishable in arrival time, frequency and spatial degree of freedom, and the postselection on two polarization events results in a 4-photon GHZ state. By choosing the setting angle of the half-wave plate to be $H_0=\theta/2$, the two photon pairs are prepared into a family of entangled states $|\phi\rangle=\cos\theta|HH\rangle+\sin\theta|VV\rangle$. The state $|\psi\rangle$ can be obtained when one of the photons acts as a trigger and a phaser properly adjusts the relative phase between $|HHH\rangle$ and $|VVV\rangle$.

In the measurement module, each of the three parties receives one photon, and we employ waveplates and polarizing beam splitter (PBS) to conduct single-qubit measurements according to their inputs. Here we choose the measurements in the form Eq.~(\ref{M}) to achieve the maximum violation of FGI. For Alice’s measurement, we fix $A_0=\sigma_x$ and $A_1=\sigma_y$. The former can be realized by a HWP with $H_{A_0}=22.5^\circ$, a PBS and two single-photon avalanche photodiodes (APDs), whereas the latter can be realized by changing the HWP to a combination of sandwich type QWP-HWP-QWP sequence with $\{Q^1_{A_1}=0, H_{A_1}=\pi/4, Q^2_{A_1}=\pi/2\}$. Two outcomes of $a_i$ are directly read by APDs ($D_1$ and $D_2$). Since Charlie’s measurements in Eq.~(\ref{M}) are as same as Alice, we set the waveplates angles as $H_{C_0}=22.5^\circ$ and $\{Q^1_{C_1}=0, H_{C_1}=\pi/4, Q^2_{C_1}=\pi/2\}$ to realize $C_0$ and $C_1$ respectively. Two outcomes of $c_k$ are directly read by APDs ($D_5$ and $D_6$). Bob's measurements are related to the state parameter $\theta$, so the angles of waveplates change with the variation of the GGHZ states. Table~\ref{table1} gives the setting angles of QWP-HWP-QWP sequence to realize the measurements of Bob $B_0$ and $B_1$ with different $\theta$. Two outcomes of $b_j$ are directly read by the following APDs ($D_3$ and $D_4$).

\begin{table}
\centering
\begin{tabular}{c|ccc|ccc}
\hline\hline
\textbf{$\theta$}  &   & \textbf{$B_0$}  & 	& \textbf{$B_1$}  & \\ \hline
& \textbf{$Q^1_{B_0}$}	& \textbf{$H_{B_0}$}  & \textbf{$Q^2_{B_0}$}  & \textbf{$Q^1_{B_1}$}	& \textbf{$H_{B_1}$}  & \textbf{$Q^2_{B_1}$}\\  \hline
$0.05\pi$   & -25.9$^\circ$			& 9$^\circ$          & -46.1$^\circ$        & 0$^\circ$	   & 9$^\circ$	  & -90$^\circ$	  \\
$0.1\pi$    & -8.4$^\circ$			& 18$^\circ$         & -45.6$^\circ$        & 0$^\circ$	   & 18$^\circ$	  & -90$^\circ$	  \\
$0.15\pi$   & -26.3$^\circ$			& 27$^\circ$          & -9.7$^\circ$        & 0$^\circ$	   & 27$^\circ$	  & -90$^\circ$	 \\
$0.2\pi$     & -5$^\circ$			& 36$^\circ$          & -13	$^\circ$        & 0$^\circ$	   & 36$^\circ$	  & -90$^\circ$	 \\
$0.25\pi$    & 0$^\circ$			& 45$^\circ$         & 0	$^\circ$        &0$^\circ$	   & 45$^\circ$	  & 90$^\circ$	 \\
$0.3\pi$    & -2.4$^\circ$			& -36$^\circ$      & 20.4	$^\circ$        & 90$^\circ$	   & 36$^\circ$	  & 0$^\circ$	 \\
$0.35\pi$    & 18.9$^\circ$			& -27$^\circ$      & 17.1	$^\circ$        & 90$^\circ$	   & 27$^\circ$	  & 0$^\circ$	 \\
$0.4\pi$    & 39.5$^\circ$			& -18$^\circ$      & 14.5	$^\circ$        & 90$^\circ$	   & 18$^\circ$	  & 0$^\circ$	 \\
$0.45\pi$    & -33.5$^\circ$		& -9$^\circ$      & -74.5	$^\circ$        & 90$^\circ$	   & 9$^\circ$	  & 0$^\circ$	 \\ \hline\hline
\end{tabular}
\caption{The different setting angles of the sandwich type QWP-HWP-QWP sequence to realize the measurement operators of Bob to achieve the maximum violation of FGI.}
\label{table1}
\end{table}

For the photon detection, we register the coincidence rates between APDs of Alice, Bob, Charlie and the trigger. Define the coincidence counting photon numbers of the joint measurement with the results $\{a, b, c\}$ are $N\{D_l, D_m, D_n, D_7\}$, where $l\in\{0,1\}, m\in\{3,4\}, n\in\{5,6\}$. For instance, the coincidence counting $\{D_1, D_3, D_5, D_7\}$ represents the result $a=b=c=0$. The conditional probabilities satisfy $P(c|ab)=P(abc)/P(ab)$, thus, the violation $S$ in Eq.~(\ref{FGI}) can be calculated by the raw coincidence counting data. Here we show the method to calculate the conditional probability $P(0_{C_0}|1_{A_0}1_{B_0})$ as an example, that is
\begin{equation}
\begin{aligned}
P(0_{C_0}|1_{A_0}1_{B_0})&=\frac{P(1_{A_0}1_{B_0}0_{C_0})}{P(1_{A_0}1_{B_0})}   \\
&=\frac{N\{D2, D4, D5, D7\}/N_{\text{total}}}{N\{D2, D4, D5, D7\}+N\{D2, D4, D6, D7\}/N_{\text{total}}}    \\
&=\frac{N\{D2, D4, D5, D7\}}{N\{D2, D4, D5, D7\}+N\{D2, D4, D6, D7\}},
\end{aligned}
\end{equation}
where $N_{\text{total}}=\sum_{l,m,n}\{D_l, D_m, D_n\}$ is the total coincidence counting photon numbers during once measurement process.

\section{Conclusion and discussion}\label{sec:4}
In the present work, we present an experimental proposal to demonstrate the FGSI in the 2SDI scenario for any three-qubit GGHZ states. In this proposal, we give a method to prepare the three-qubit GGHZ states with different state parameters in photon system. The measurement observalbes in each party can be realized by different waveplates and a PBS. By choosing the angles of the sandwich type QWP-HWP-QWP sequence, our experiment proposal can observe the maximum quantum violation for any three-qubit GGHZ states. The proposal can be easily extended to high-dimensional qubits and multi-photon GHZ states, allowing for the entanglement of multi-party systems, which provides a method to study the complex multi-party quantum protocols.

We have shown that the maximum quantum violation of the tripartite steering inequality for pure three-qubit GGHZ states. This result brings to mind the concept of quantum self-test since any pure GGHZ state $|\psi\rangle$ violates Eq.~(\ref{FGS}) to its algebraic maximum of 4. Self-test is an important concept in quantum information theory, referring to the process of verifying the state or operation of a quantum system using experimental data or measurement outcomes, without needing to fully trust the experimental device or control systems. However, in Ref.~\cite{gupta2022all}, the authors prove that the maximal violation of FGI certifies genuine entanglement in three-qubit pure states in the 2SDI scenario, not only the GGHZ states. They show a part of generated pure W-class states can maximally violate the FGI numerically. Thus, self-test to GGHZ states is still an open question.

\begin{acknowledgements}
Z.H.B. was supported by the National Natural Science Foundation of China (Grant No. 12104186). Y.S. was supported by the NNSF of China (Grant No. 12401597), the NSF of Jiangsu Province (Grant No. BK20241603), and the Wuxi Science and Technology Development Fund Project (Grant No. K20231008).
\end{acknowledgements}

\bibliographystyle{plain}
\bibliographystyle{apsrev4-1}
\bibliography{reference.bib}

\end{document}